\documentclass{snmp2001}

\begin{document}
\setcounter{page}{540}%

\FirstPageHeading{author}

\ShortArticleName{Quantum Algebras, Particle Phenomenology, 
                          and (Quasi)Supersymmetry} 

\ArticleName{Quantum Algebras, Particle Phenomenology,   \\
            and (Quasi)Supersymmetry}

\Author{A. M. GAVRILIK} 
\AuthorNameForHeading{A. M. Gavrilik}
\AuthorNameForContents{GAVRILIK A. M.}

\Address{ Bogolyubov Institute for Theoretical Physics, 03143 Kiev,
          Ukraine}
\EmailD{omgavr@bitp.kiev.ua}

\Abstract{Quantum algebras $U_q({\rm su}_n)$ used as the algebras 
of flavour symmetry (usually described by $SU(n)$) to study static 
properties of hadrons lead to intriguing results.
In this contribution we focus on the peculiar properties 
manifested by different $q$-deformed structures (e.g., 
the braided line, the quantum algebras $U_q({\rm su}_2)$ and 
$U_q({\rm su}_n)$, $\ n\ge 3$) in the special limit of $q=-1$.  
Similarities (complete or partial) with supersymmetry 
that emerge in this special limit are discussed. }

\section{Introduction}

Our goal is to pay special attention to the exotic situation 
that arises if, within the application of 
quantum algebras $U_q({\rm su}_n)$       \cite{gavr:drin,gavr:klim}
to phenomenological description (see     \cite{gavr:nucl,gavr:nato} 
and refs. therein) of basic static properties of hadrons - 
vector mesons as well as baryons, one restricts itself 
to the peculiar case $q=-1$ of the deformation parameter.
In the paper, we first briefly mention the two more or less 
realistic appearances of supersymmetry (SUSY) algebras 
applied directly in the sector of hadron mass spectrum. 
Note that the first appearance of SUSY in the context of 
hadron physics goes back to Miyazawa's paper    \cite{gavr:miya}.
It employs a kind of superalgebra which is connected 
with internal symmetry and extends the usual $SU(3)$ scheme 
by means of baryon number changing currents.
In that paper, the author has suceeded to derive,
based on a superalgebra, the mass sum rules other than
the celebrated Gell-Mann - Okubo (GMO) one, that is,
$m_{N}+m_{\tiny{\Xi}} = 
\frac32 m_{\tiny \Lambda} + \frac12 m_{\tiny \Sigma}$.
On the contrary, the spectrum denerating 
(or dynamical) superalgebra used in       \cite{gavr:bohm}
incorporated a superization of space-time symmetry and
gave a possiblity to analyse the towers of excited states,
for each ground state baryon (e.g., nucleon) or vector
meson (e.g., $\rho$-meson). 
We discuss these two examples in Section 2.
Then, Sections 3 and 4 are devoted to the very instructive 
examples of $q$-deformed structure which, if one sends 
$q\to -1$, show either exact SUSY (the case of braided line 
whose relation to SUSY is considered in Section 3), or 
the features only reminiscent of supersymmetry, see Section 4. 
In the 5th section we deal with the peculiar case of $q= -1$ 
concerning the quantum algebras $U_q({\rm su}_n)$ which appear 
in the context of their use as the algebras describing 
flavor symmetries of hadrons and enabling to derive new, 
very precise mass relations. 
In this scheme, the restriction to the limit $q= -1$ is 
physically motivated.

\section{Dynamical supersymmetry and hadron mass spectrum}

In  \cite{gavr:miya}    
the two copies of superalgebra, namely, 
\begin{gather}
[F_i,F_j]={\rm i} f_{ijk} F_k ,  \hspace{10mm} 
[F_i,G_j]={\rm i} f_{ijk} G_k ,  \hspace{10mm} 
\{G_i,G_j\}= d_{ijk} F_k ,           \\    \nonumber        
[\bar{F}_i,\bar{F_j}]={\rm i} f_{ijk} \bar{F}_k , 
        \hspace{10mm}         
[\bar{F}_i,\bar{G_j}]={\rm i} f_{ijk} \bar{G}_k ,      
        \hspace{10mm}  
 \{\bar{G}_i,\bar{G_j}\}= - d_{ijk} \bar{F}_k ,
\end{gather}
have been introduced.  
For their realization, the conventional $3\times 3$
 hermitian matrices $\lambda_i$ ($i=0,1,2,3,8$ for the $F_i,\ \bar{F}_i$, 
and $i=4,5,6,7$ for the $G_i,\ \bar{G}_i$) have been utilized.
By means of symmetry breaking terms ($C^{3b}_{3b}$, 
$C^{a3}_{a3}$ and $C^{33}_{33}$ in the notation of  \cite{gavr:miya}) 
which provide mass splitting between quarks and 
diquarks (i.e., SUSY breaking), as well as splitting
between isomultilplets (breaking of $SU(3)$ to $SU(2)$), 
instead of the standard GMO mass relation the formulas 
\begin{equation}
m_{\tiny N}+m_{\tiny \Xi} = 
m_{\tiny \Lambda} + m_{\tiny \Sigma}      \hspace{10mm}   
m_{\tiny Y_0^*} = m_{\tiny \Sigma}        \hspace{10mm}   
2 m_{\tiny K^*} = m_{\rho} + m_{\phi}     \hspace{10mm}   
 m_{\rho} = m_{\omega} 
\end{equation}
for baryons and for vector mesons have beeen obtained.

A completely different scheme for treating 
hadron mass spectrum developed in                \cite{gavr:bohm}
employs a particular {\em dynamical superalgebra} 
${\rm Osp}(1|4)$ connected with space-time symmetries.  
The dynamical superalgebra with generators 
$S_{\mu\nu},\Gamma_\mu,Q_\alpha$ respects the chain  
\[
{\rm Osp}(1|4)_{\tiny S_{\mu\nu},\Gamma_\mu,Q_\alpha}\supset 
{\rm SO}(3,2)_{S_{\mu\nu},\Gamma_\mu}\supset
{\rm SO}(3,1)_{S_{\mu\nu}}
\]
where for the subalgebras ${\rm SO}(3,1)_{S_{\mu\nu}}$ 
and ${\rm SO}(3,2)_{S_{\mu\nu},\Gamma_\mu}$ the generators  
$S_{\mu\nu}$ and $\Gamma_\mu$ obey
\begin{gather}
[S_{\mu\nu},S_{\rho\sigma}]=-i
(\eta_{\mu\rho} S_{\nu\sigma} + \eta_{\nu\sigma} S_{\mu\rho}  
- \eta_{\nu\rho} S_{\mu\sigma}-\eta_{\mu\sigma} S_{\nu\rho})\ , \\ 
[S_{\mu\nu},\Gamma_{\rho}]=-i
(\eta_{\mu\rho} \Gamma_{\nu} - \eta_{\nu\rho} \Gamma_{\mu})\ ,
              \hspace{10mm}
[\Gamma_{\mu},\Gamma_{\nu}] = -i S_{\mu\nu}\ .
\end{gather}
The relations involving anticommuting charges $Q_\alpha$ 
and $\bar Q_\beta=-(Q^TC)_\beta$, namely
\begin{gather}
[S_{\mu\nu},Q_\alpha] = 
-\frac12 \left(\sigma^s_{\mu\nu}\right)_{\alpha}^{\ \beta} Q_\beta \ , 
                 \hspace{10mm}
[\Gamma_\mu, Q_\alpha] =
-\frac12 \left(\gamma_{\mu}\right)_{\alpha}^{\ \beta} Q_\beta \ , 
                                                          \nonumber  \\
\{ Q_\alpha , {\bar Q}_\beta \} =
-\frac12 \left(\sigma^{\mu\nu}\right)_{\alpha\beta} S_{\mu\nu}
+ \left(\gamma^{\mu}\right)_{\alpha\beta} \Gamma_\mu \ ,    \nonumber
\end{gather}
along with (3),(4), complete the symmetry algebra 
to the superalgebra 
${\rm Osp}(1|4)_{\tiny S_{\mu\nu},\Gamma_\mu,Q_\alpha}.$ 

To construct the hamiltonian, supercharges should be 
incorporated (like in supersymmetric quantum mechanics), 
through the term 
$\frac{1}{2n}\sum_{\alpha=1}^n \{ Q_\alpha , Q_\alpha^\dagger \}$.
The resulting Hamiltonian
\[
H=v\Biggl( P_\mu P^\mu -\frac{1}{\alpha'}\frac14 
\sum_{\beta=1}^4 
\{ Q_\beta , Q_\beta^\dagger \} - \tilde m_0^2\Biggr)  
\]
is to be completed by Casimirs of subalgebras in
the chain
$ {\rm Osp}(1|4)\supset 
{\rm SO}(3,2)_{S_{\mu\nu},\Gamma_\mu}\supset
{\rm SO}(3)_{S_{ij}}$  $\times \ {\rm SO}(2)_{\Gamma_0}.$
In its final form, the Hamiltonian reads  
\begin{equation}
H=v\Bigl( P_\mu P^\mu -
\frac{1}{\alpha'}\hat P_\mu \Gamma^\mu
- \lambda^2 \hat W 
+ \beta \hat C_{SO(3,2)} - \tilde m_0^2\Bigr) 
\end{equation}
and, correspondingly, hadron 
mass spectrum is described by the formula          \cite{gavr:bohm}
\begin{equation}
m^2 =  -\frac{1}{\alpha'} \mu
+ \lambda^2 j(j+1)+\beta (2-2s^2) + \tilde m_0^2 \ .
\end{equation}
In this expression, 
$1/\alpha'$  (related to the slope of Regge trajectory),
$\lambda^2$, and $\beta$ are empirical system parameters; 
$\mu$ resp. $j(j+1)$ are eigenvalues of 
$\hat P_\mu \Gamma^\mu$ resp. $\hat W$;
$s$ labels $SO(3,2)$ representations, and
$\tilde m_0$ is the background mass.

Comparison of the mass formula (6) with experimental data, 
using the particular representation 
$D(\frac32,\frac12)\ \oplus \ D(2,1)$
of the dynamical superalgebra, shows that 
the series (tower) of excited states over 
the lowest lying $1^-$ vector mesons $\rho$ or $\omega$ 
and the $\frac12^+$ nucleon's tower (its resonances)
fit the data very well if one sets:
$\frac{1}{\alpha'}({\rm meson})\sim \frac{1}{\alpha'}({\rm nucleon})$
and $\lambda^2({\rm meson})\sim \lambda^2({\rm nucleon})$. 
It is this fact that was interpreted in       \cite{gavr:bohm}
as empirical evidence for supersymmetry in 
the hadron mass spectra.
This observation may be considered as an 
extension of the well-known success of 
dynamical supersymmetries in nuclear physics      \cite{gavr:bars}  
to the level of hadrons. 

\section{$q$-Deformed oscillator at $q \to -1$ and supersymmetry}

In   \cite{gavr:azca}  it was shown that the $q$-deformed 
calculus on the braided line               \cite{gavr:maji} 
(tightly connected with $q$-deformed oscillator),  
in the nontrivial particular case of $q=-1$ exhibits     
supersymmetric properties. In this section we discuss 
some details of this correspondence, following    \cite{gavr:azca}.
 
The braided (or $q$-deformed) line is defined  \cite{gavr:maji}
in terms of a single non-commuting variable $\theta$ which obeys 
a Hopf algebra structure operating with coproduct,
\begin{gather}
\Delta\theta = \theta\otimes 1 + 1\otimes\theta,     \\
(1\otimes\theta)(\theta\otimes 1) =  q\theta\otimes\theta ,
                      \hspace{18mm}
(\theta\otimes 1)(1\otimes\theta) =  \theta\otimes\theta .
\end{gather}
as well as a counit and antipode. Note that it is the first 
relation in (8) that determines the nontrivial (for $q\ne 1$) 
braiding. 

With $[X,Y]_z\equiv XY-zYX$, denoting $\theta=1\otimes\theta$  
and $\delta\theta=\epsilon=\theta\otimes 1$  
as in        Ref. \cite{gavr:maji},  
yields
\[ 
[\epsilon,\theta]_{q^{-1}} = 0          \hspace{18mm}
                              {\rm and}  \hspace{18mm}
\Delta\theta = \epsilon + \theta .
\]  
Here the latter equality corresponds to (7);  
it encodes the action upon $\theta$ of the left 
translation by $\epsilon$, 
$L_{\epsilon} \theta:$ $\ \theta \mapsto\epsilon + \theta$.
As seen, $\epsilon$ and $\theta$ anticommute when $q=-1$.

To construct a differential calculus on the braided line,
one introduces a left derivation operator with 
respect to $\theta$, obeying 
$[\epsilon{\cal D}_L,\theta]=\epsilon$, so that
\begin{equation}
[{\cal D}_L,\theta]= 1 ,  \hspace{18mm} 
                              \frac{d}{d\theta}\theta = 1 .
\end{equation}
Likewise, one can introduce right shifts $R_{\eta}\theta :$  
$\ \theta\mapsto \theta + \eta$ by odd parameter $\eta$ so that
$[\theta,\eta]_{q^{-1}}=[\eta,\theta]_q=0$ 
(again, $\theta$ and $\eta$ anticommute if $q=-1$). 
The right derivative operator satisfies
$[\theta,{\cal D}_R]= 1 $ and also the relation
\begin{equation}
{\cal D}_R = -q^{-(1+N)} {\cal D}_L 
\end{equation}
involving the number operator $N$ defined according to
\begin{equation}
[N,\theta]= \theta ,  \hspace{18mm} 
                      [N,{\cal D}_L]= - {\cal D}_L \ .
\end{equation}
The differential calculus defined by (9)-(11) 
at generic $q$ is called $q$-calculus.

With the identification
$\theta = a^\dagger ,  \hspace{3mm} 
                      {\cal D}_L = q^{N/2} a \ ,
$
the $q$-calculus is related to the 
$q$-deformed harmonic oscillator           \cite{gavr:bied}
\begin{equation}
a a^\dagger - q^{\mp 1/2} a^\dagger a = q^{\pm N/2}\ .
\end{equation}
The entity $q^{1/2}$ and its  
power $(q^{1/2})^N$ in (12)
are of importance since, from (12), by 
exploiting hermitian conjugacy one comes to 
the formulas  $a a^\dag = [N+1]_{q^{1/2}}$ 
and $a^\dag a = [N]_{q^{1/2}}$ valid 
for the $q$-deformed oscillator           \cite{gavr:bied}
of Biedenharn and Macfarlane. 
Here $[A]_z\equiv(z^A-z^{-A})/(z-z^{-1})$.

Let $\lfloor A\rfloor_q\equiv(1-q^A)/(1-q)$. 
A function of $\theta$ given by the expansion 
$f(\theta)=\sum^\infty_{m=0}C_m\theta^m/\lfloor m\rfloor_q!$ 
admits the derivative $\frac{d}{d\theta}f(\theta)=
\sum^\infty_{m=0}C_m{\theta}^m/\lfloor m-1\rfloor_q!$
implying  that
\[
\left[{\cal D}_L,\frac{\theta^m}{\lfloor m\rfloor_q!}\right]_{q^m}=
\frac{\theta^{m-1}}{\lfloor m-1\rfloor_q!} \ .               
\]
The difficulties appearing in the limit $q\to-1$ already 
at $m=2$ (since $\lfloor 2\rfloor_q=0$ in this limit) are 
tamed by setting $q=-1+iy$ and letting $y\to 0$. 
Then, the definition
\begin{equation}
t := \lim_{\tiny{q\to -1}}    
          ({i\theta^2}/{\lfloor 2\rfloor_q!}) 
\end{equation}
implying that, with $\theta^2=0$ imposed, the limit of 
the ratio in (13) should be finite and nonzero, imports
the additional variable $t$ as a necessary ingredient 
of the braided line if $q\to-1$. As shown in       \cite{gavr:azca},
in this limit the terms of the form 
$\theta^{2r+p}/\lfloor 2r+p\rfloor_q!$
also can be handled by means of $t$. 
Due to this, any function $f(\theta)$ on the braided line 
(generic $q$), reduces in the limit $q\to-1$ to a 'superfield'
given by the function $f(t,\theta)$. 

It can be shown that $[{\cal D}_L^2,t]=i$ and, 
with the definition
\[
\{ {\cal D}_L,{\cal D}_L \} = 2i\partial_t
\hspace{18mm}  {\rm or}  \hspace{18mm}
\partial_t =-i{\cal D}_L^2 \ ,
\]
the relation $[\partial_t,t]=1$ is valid.
The operator ${\cal D}_L$ then becomes the supercharge, 
${\cal D}_L\equiv Q$, of one-dimensional supersymmetry, 
and one comes to the relations: 
\[
Q=\partial_\theta + i\theta\partial_t \ ,
    \hspace{18mm}      \{ Q , Q \} = 2i\partial_t \ .
\]
Likewise, the operator $D={\cal D}_R=(-1)^N{\cal D}_L$
becomes the (super)covariant derivative so that  
\[
D=\partial_\theta - i\theta\partial_t \ ,
          \hspace{18mm}  \{ D , D \} = -2i\partial_t \ ,
\hspace{14mm}  {\rm and} \hspace{14mm}  \{ Q , D \} = 0 .
\]
Another interesting result derived in      \cite{gavr:azca}
is the coproduct for $t$ with unusual $\theta$-dependent term:
\[
\Delta t = t\otimes 1 + 1\otimes t + i\theta\otimes\theta .                      
\]

Thus, proper treatment of braided line in the 
peculiar limit $q\to -1$ shows that, in this limit, 
an additional variable $t$ related to $\theta^2$ (see (13)), 
as well as to higher powers, must arise      \cite{gavr:azca}.
As a result, the braided line at $q\to -1$ is made up 
of the two variables $\theta$ and $t$ which span the 
one-dimensional superspace, SUSY being the translational
invariance along this line.

\section{Example of Zachos, based on the $q=-1$ limit of   
         $U_q({\rm su}_2)$ }

Quantum algebra $U_q({\rm su}_2)$         \cite{gavr:drin,gavr:klim}
is generated by the elements $I_+, I_-, I_0,$
obeying the relations   
\begin{gather}
[I_0,I_{\pm}] = \pm I_{\pm}\ ,   \hspace{14mm}
[I_+, I_-] = [2 J_0]_q \equiv 
            ({q^{2 J_0}-q^{-2 J_0}})/({q-q^{-1}})\ , 
                                                        \nonumber\\
\Delta(J_0) = J_0\otimes 1 + 1 \otimes I_0\ , \hspace{12mm}
\Delta(J_{\pm}) = 
J_{\pm}\otimes q^{-J_0} + q^{+J_0}\otimes J_{\pm}\ 
\end{gather}
and the relations that involve antipode 
and counit (which will not be used here). 

As shown in      \cite{gavr:zach},  
this quantum algebra exhibits an intriguing features at
the level of its representations when the deformation 
parameter $\ q=-1$. Let us consider this example. 

Using coproduct, one can form composites of 
two spin $\frac12$ doublets according to 
${\bf 2}\otimes {\bf 2} = {\bf 3} \oplus {\bf 1}$:
\[
{\rm singlet} 
        \hspace{10mm}    \longleftrightarrow    \hspace{14mm}
\alpha = |q^{1/2} \uparrow\downarrow 
         - q^{-1/2} \downarrow\uparrow\rangle         
\]
 \vspace{-8mm}
\[
 {\rm triplet} \hspace{6mm} \longleftrightarrow \hspace{6mm}
\left\{
\begin{array}{ll}
             {} &  \beta = |\uparrow\uparrow\rangle\ ,  \\  
             {} &  \Delta(J_-)\beta = 
                  \frac{1}{\sqrt{2}}|q^{1/2}\uparrow\downarrow 
                         + q^{-1/2}\downarrow\uparrow\rangle\ ,  \\ 
             {} & (\Delta(J_-))^2\beta = |\downarrow\downarrow\rangle  
\end{array}  
\right.                                                
\]
For $q=1$, the singlet state is antisymmetric 
whereas each of the triplet states is symmetric.
Now let $q=-1$. In this case the multiplets turn into
\[             \hspace{2mm}
\alpha = | i \uparrow\downarrow 
          - \frac{1}{i} \downarrow\uparrow\rangle  
                     \hspace{33mm}     ({\rm symmetric})        
\]
\vspace{-8mm}
\begin{equation}   
\begin{array}{ll}
   \beta = |\uparrow\uparrow\rangle\ , 
                      \hspace{12mm}    & ({\rm symmetric} ) \\                                                       
         \Delta(J_-)\beta = 
               \frac{1}{\sqrt{2}}| i \uparrow\downarrow 
                + \frac{1}{i}\downarrow\uparrow\rangle\ ,   
                \hspace{12mm}      &  ({\rm antisymmetric})  \\ 
(\Delta(J_-))^2\beta = |\downarrow\downarrow\rangle\ . 
                   \hspace{12mm}    &    ( {\rm symmetric} )                     
\end{array}  
\end{equation}
It is seen from (15) that the coproduct operation 
$\Delta(J_-)$ changes the symmetry of wave function. 
That is, rasing and lowering operators in the coproduct 
act as statistics-altering operators.
Although the constituents of the states haven't been 
converted to fermions, this alteration of the symmetry 
of wave function {\em is reminiscent of} SUSY. 
It is instructive to compare
this structure with $N=2$ supersymmetric quantum mechanics, 
stressing both similarities and peculiar features.

Consider (graded) direct product of two copies of SUSY QM algebras:
\begin{gather}
S S^\dag + S^\dag S = 1\ ,        \hspace{8mm}   
s s^\dag + s^\dag s = 1\ ,        \hspace{8mm} 
S^\dag S^\dag = s^\dag s^\dag = S S = s s = 0\ ,  \nonumber \\
s S + S s = 0\ ,                     \hspace{8mm} 
s^\dagger S^\dagger + S^\dagger s^\dagger = 0\ ,   
                                     \hspace{8mm} 
s S^\dagger + S^\dagger s = 0\ ,     \hspace{8mm}  
s^\dagger S + S s^\dagger = 0\ .                    
\end{gather}
This graded Lie algebra can be obtained, using 
appropriate Wigner-Inon\"{u} contraction, from the simple 
Lie superalgebra $SU(2|1)$ (realizable in terms of 
Gell-Mann $SU(3)$ $\lambda$-matrices so that 
$\{ \lambda_1, \lambda_2, \lambda_3, \lambda_8\} $ constitute
even generators whereas 
$\{ \lambda_4, \lambda_5, \lambda_6, \lambda_7\} $ constitute
odd generators).

One can realize the algebra (16) on two boson states 
$|B\rangle$, $|b\rangle$, and two fermion states
$|F\rangle$, $|f\rangle$, as:
$ S~|B\rangle = |F\rangle,\ \ s |b\rangle = |f\rangle,\ \
S^\dagger |F\rangle = |B\rangle,\ \ s^\dagger |f\rangle = |b\rangle .$                                         
The (nullifying) rest of actions reads:  
$ S~|F\rangle = S |b\rangle = s |B\rangle = s^\dagger |F\rangle =
s^\dagger |b\rangle = S^\dagger |f\rangle = 
S^\dagger |B\rangle = s|f\rangle = 0 . $          
With their use,
\begin{equation}
s |Bb + bB\rangle =  |Bf + fB\rangle\ ,    \hspace{18mm}
Ss |Bb + bB\rangle =  |Ff - fF\rangle\ .            
\end{equation}
Thus, $\Delta(J_-)$ in (15) switches the symmetry of 
wave function like the even (bosonic) operator $Ss = - sS$, 
see (17), but only the latter is nilpotent 
due to nilpotency  of $S , s$.
The other important difference consists in the structure and 
dimensionality of multiplets. Namely, for $q=-1$ these remain
the same as in the classical case of $su(2)$ Lie algebra. 
On the other hand, for graded Lie algebra 
the representations are of different dimensions 
(compare, e.g., $SU(2|1)$ and $SU(3)$).
Hence, the conclusion: this $q=-1$ case implies a kind of 
{\em quasi-supersymmetry}.

\section{GMO formula and $U_q(su_n)$ at $q=-1$}

One can either utilize representation-theoretic aspects 
of the quantum algebra $U_q(su_n)$ or,
alternatively, construct the mass operator using 
$q$-tensor operators. In the latter case        \cite{gavr:acco}, 
main   
ingredients of the Hopf algebra structure of $U_q(su_n)$ 
(comultiplication $\Delta$ and antipode $S$) play the role .  
The $\Delta$ and $S$ are defined            \cite{gavr:drin,gavr:klim} 
on the $U_q(su_n)$ generators
$E_i^{\pm}$ and  $H_i$  as 
\begin{gather}
S(E_i^\pm)=-q^{H_i/2}E_i^\pm \ ,\ \ \ \ \ \
S(H_i)=-H_i\ , \ \ \ \
S(q^{H_i/2})=q^{-H_i/2} \ , \ \ \ \ S(1)=1 \ ,   \nonumber  \\
\Delta(E_i^\pm)=E_i^\pm\otimes q^{H_i/2} +
q^{-H_i/2}\otimes E_i^\pm \ , \ \ \ \ \ \ 
\Delta (H_i)= H_i\otimes 1+1\otimes H_i q^{-H_i/2}.
\end{gather}
The adjoint action of $U_q(su_n)$ defined     \cite{gavr:klim}
as 
$ad_A B = \sum { A_{(1)} B S(A_{(2)})}$
with 
$A,B \in U_q(su_n)$ and $A_{(1)},$ $\ A_{(2)}$ 
determined from 
$ \Delta (A) = \sum { A_{(1)} \otimes A_{(2)}},$
with the account of (18) reads:
\begin{gather}
ad_{H_i} B=H_i B 1+1 B S(H_i)=H_i B-B H_i\ ,       \nonumber   \\
ad_{E_i^\pm}B=
E_i^\pm B q^{-H_i/2}-q^{-H_i/2}B q^{H_i/2}E_i^\pm q^{-H_i/2}\ .\nonumber
\end{gather}
The $q$-tensor operators                     \cite{gavr:kli2} 
transforming under the adjoint action 
of $U_q(su_3)$ as ${{\bf 3}}$ and ${{\bf 3}^*}$, 
consist of the triples $(V_1,V_2,V_3)$ and 
$(V_{\bar{1}},V_{\bar{2}},V_{\bar{3}}),$ 
respectively.
Let $[X,Y]_q\equiv XY-qYX$. 
It can be shown that the particular triple   
of elements from $U_q(su_4)$
\begin{gather}        
V_1=[E_1^+,[E_2^+,E_3^+]_q]_q\ q^{-H_1/3-H_2/6} ,  \nonumber   \\
V_2=[E_2^+,E_3^+]_q\ q^{H_1/6-H_2/6} ,             \hspace{8mm} 
V_3=E_3^+ q^{H_1/6+H_2/3}                              
\end{gather}
transform as ${{\bf 3}}$ under $U_q(su_3),$
$V_1$ corresponds to the highest weight vector, the pair
$(V_1,V_2)$ is $U_q(su_2)$ (iso)doublet and $V_3$ its singlet. 
Likewise one constructs from elements of $U_q(su_4)$ 
the triple $(V_{\bar{1}},V_{\bar{2}},V_{\bar{3}})$
that transforms as ${{\bf 3}}^*$ under adjoint action of
$U_q(su_3),$  where $V_{\bar{3}}$ corresponds to the 
highest weight vector, the pair $(V_{\bar{1}},V_{\bar{2}})$ 
is isodoublet and $V_{\bar{3}}$ is $U_q(su_2)$ singlet.

The mass operator $\hat M = \hat M_0 + \hat M_8 $
involves $\hat{M}_0$, as $U_q(su_3)$ scalar, 
and the term $\hat{M}_8$ transforming as 
the $I=0,Y=0$ component of tensor operator of 
${\bf 8}$-irrep of $U_q(su_3).$
The irrep ${\bf 8}$ occurs twice 
in the decomposition
$ {{\bf 8}} \otimes {{\bf 8}} =
  {{\bf 1}} \oplus {{\bf 8}}^{(1)} \oplus
  {{\bf 8}}^{(2)} \oplus
  {{\bf 10}}^* \oplus {{\bf 10}} \oplus
  {{\bf 27}} .
$
Then, usage of Wigner-Eckart theorem for 
$U_q(su_n)$ quantum algebras                    \cite{gavr:kli2}
applied to $q$-tensor operators transforming 
as irrep ${\bf 8}$ of $U_q(su_3)$, turns 
the mass operator into  
$\hat M = \hat M_0 + \hat M_8 = 
M_0 {\bf 1}+\alpha V_8^{(1)}+\beta V_8^{(2)}.$
Here  ${\bf 1}$ is the identity operator,
$V_8^{(1)}$ and $V_8^{(2)}$ are two fixed 
tensor operators with non-proportional matrix elements,
each transforming as the $I=0, Y=0$ component 
of irrep ${\bf 8}$ of $U_q(su_3)$;   
$\ M_0,$ $\alpha$ and $\beta$ are some  
constants depending on details (dynamics) of the model.

If $|B_i\rangle$ is a basis vector of representation
${\bf 8}$ space which corresponds to some $(1/2)^+$ baryon, 
then the mass of this baryon is calculated as
\begin{equation}
M_{B_i} = \langle B_i|\hat{M}|B_i \rangle =
\langle B_i |(M_0 {\bf 1} + \alpha V_8^{(1)} + 
\beta V_8^{(2)})| B_i \rangle\ .
\end{equation}

The decompositions
$ {{\bf 3}} \otimes {{\bf 3}}^* =
  {{\bf 1}} \oplus {{\bf 8}},\ \
  {{\bf 3}}^* \otimes {{\bf 3}} =
  {{\bf 1}} \oplus {{\bf 8}}                
$
imply that the operators $V_{3}\ V_{\bar{3}}$ and 
$V_{\bar{3}}\ V_{3}$ formed from $V_{3}$ in (19) 
and $V_{\bar{3}}$   
are just the two isosinglets $V_8^{(1)}, \ V_8^{(2)}$ 
needed in (20).   Hence, the mass operator in (20) 
can be rewritten (redefining $M_0,\alpha,\beta$) 
in the equivalent form
\begin{equation}
\hat M = M_0 {\bf 1} +\alpha V_3 V_{\bar{3}}
+ \beta  V_{\bar{3}} V_3 =
\hat M = M_0 {\bf 1} +\alpha E_3^+ E_3^- q^Y + \beta E_3^- E_3^+ q^Y
\end{equation}
where the hypercharge $Y=(H_1+2H_2)/3$ has been introduced.

To calculate matrix elements (20) using (21)  
we embed the octet ${\bf 8}$ in a particular irrep 
of $U_q(su_4);$ embedding it, e.g., in ${\bf 15}$ (adjoint) 
irrep of $U_q(su_4)$, we get the octet baryon masses  
\begin{equation}
 M_N=M_0+\beta q \ , \ \ \     M_\Sigma= M_0\ ,
\ \ M_\Lambda =M_0 + [2]_q[3]_q^{-1} (\alpha+\beta)\ ,
\ \ M_\Xi=M_0+\alpha q^{-1}  
\end{equation}
(obviously, the expressions for $M_N,\ M_\Xi$ are
not invariant under $q \to q^{-1}$).  Excluding  
$M_0,\ \alpha$ and $\beta$ from (22) results in the 
following $q$-analogue of GMO formula for octet baryons:
\begin{equation}
[3]_qM_\Lambda + M_\Sigma=[2]_q(q^{-1} M_N+q M_\Xi)\ .
\end{equation}

Using empirical masses, the deformation parameter $q$ 
is fixed by fitting: for each of the 
$q_{1,2}=\pm 1.035$, $q_{3,4}=\pm 0.903 \sqrt{-1}$, 
the $q$-deformed mass relation (23) holds within 
experimental uncertainty (although for $q_3,\ q_4$ the 
constants $\alpha$ and $\beta$ in (22) must be pure imaginary).

The right hand side of eq. (23) is invariant under 
$q \to q^{-1}$ only if $q=q^{-1}$, that is, if $q=\pm 1.$
Behind the 'classical' GMO mass formula 
which obviously follows from (23) at $q=1$ and
corresponds to the nondeformed unitary symmetries
$SU(4) \supset SU(3) \supset SU(2)$, there is also an unusual
'hidden symmetry' reflecting the singular $q=-1$ case of
$U_q(su_4) \supset U_q(su_3) \supset U_q(su_2)$ algebras,
undefined in this case.  The relevant objects, 
however, exist as operator algebras             \cite{gavr:acco}. 
Let us describe them in the part 
corresponding to $n=2$ and $n=3$.

At generic $q$, $q\ne -1$, the algebra $U_q(su_2)$ is generated by the
elements $E^+$,$E^-$ and $H$, which satisfy the relations
\begin{equation}
[H,E^\pm]=\pm 2 E^\pm ,   \hspace{18mm}    [E^+,E^-]=[H]_q \ .
\end{equation}
In the limit $q\to 1$ it reduces to the nondeformed $su_2$.
We take the representation spaces of the latter in order 
to construct operator algebras for the case $q=-1$. 
To each  $su_2$ representation space given by $j$ 
(which takes integral or
half-integral nonnegative values) with basis elements
$|j m\rangle,\ m=-j,-j+1,\ldots,j$, there corresponds 
an operator algebra
generated by the operators defined according to the formulas
\begin{equation}
H|j m\rangle=2m|j m\rangle,\ \ \
E^+|j m\rangle=\alpha_{j,m}|j m+1\rangle,\ \ \
E^-|j m\rangle=\alpha_{j,m-1}|j m-1\rangle\ \ \
\end{equation}
where
\vspace{-0.1mm}
\[
  \alpha_{j,m}=
               \left\{
\begin{array}{ll}
             {} & \sqrt{-(j-m)(j+m+1)},\ \ \ \
                          j \ \ \ {\rm is \ an \  integer}\ , \\    
             {} &  \vspace{-0.8mm} \\ 
             {} &   \sqrt{(j-m)(j+m+1)},\ \ \ \  \
                      j \ \ \ {\rm is\ a \ half\!-\!integer} .
\end{array}  
               \right.                                                 
\]
\vspace{-0.1mm}
So defined operators $E^+$,$E^-$ and $H$ 
{\em on the basis elements} $|j m\rangle $ 
satisfy the relations (compare with (24)), 
one of which {\em depends on whether} $j$ 
{\em is an integer or a half-integer}:
\begin{equation}           
\vspace{-0.1mm}
[H,E^\pm]=\pm 2 E^\pm,\ \ \ \ \ \
[E^+,E^-]=
\left\{
\begin{array}{ll}
          {} &   -H,\ \ \ \ \ \ j \ \ \ \ {\rm is\ an\ integer}; \\
          {} & \vspace{-1.4mm} \\ 
          {} &   H,\ \ \ \ \ \ \ j \ \ \ \ {\rm is\ a \ half\!-\!integer} . 
\end{array}  
               \right.                                                 
\vspace{-0.1mm}
\end{equation}

To treat the (singular) case $q=-1$ of $U_q(su_3)$
it is more convenient to deal with $U_q(u_3)$. We take
a representation space $V_\chi$, labelled by
$\{m_{13},m_{23},m_{33}\}\equiv\chi$, of the nondeformed $u_3$
and the Gel'fand-Tsetlin basis with the basis elements
$|\chi;m_{12},m_{22};m_{11}\rangle$ in each $V_\chi$.
Define the operators $E_1^+$,$E_1^-$,$H_1$,
$E_2^+$,$E_2^-$, $H_2$ that form the operator algebra
of the $\chi$-type by their action according to the formulas
(let us denote $\sigma_{1,3}\equiv m_{11}+m_{13}+m_{23}+m_{33}$):
\begin{gather}
H_2|\chi;m_{12},m_{22};m_{11}\rangle=
(2m_{12}+2m_{22}-m_{13}-m_{23}-m_{33}-m_{11})
|\chi;m_{12},m_{22};m_{11}\rangle \ ,                     \nonumber \\
E_2^+|\chi;m_{12},m_{22};m_{11}\rangle=
a_{\chi,m_{11}}
(m_{12},m_{22})~|\chi;m_{12}+1,m_{22};m_{11}\rangle       \nonumber \\
  \hspace{38mm}     + b_{\chi,m_{11}}
(m_{12},m_{22})~|\chi;m_{12},m_{22}+1;m_{11}\rangle \ ,    \nonumber \\
E_2^-|\chi;m_{12},m_{22};m_{11}\rangle=
a_{\chi,m_{11}}
(m_{12}-1,m_{22})~|\chi;m_{12}-1,m_{22};m_{11}\rangle      \nonumber \\
  \hspace{38mm}     + b_{\chi,m_{11}}
(m_{12},m_{22}-1)~|\chi;m_{12},m_{22}-1;m_{11}\rangle       \nonumber 
\end{gather}
where
\vspace{-1mm}
\begin{gather}
\hspace{-4mm}
a_{\chi,m_{11}}(m_{12},m_{22})=
\left(
(-1)^{\sigma_{1,3}}  
\frac{(m_{13}-m_{12})(m_{23}-m_{12}\!-\!1)
(m_{33}-m_{12}\!-\!2)(m_{11}-m_{12}\!-\!1)}
{(m_{22}-m_{12}-1)(m_{22}-m_{12}-2)}
\right)^{1/2}  ,                                        \nonumber \\
\hspace{-3mm}
b_{\chi,m_{11}}(m_{12},m_{22})=
\left(
(-1)^{\sigma_{1,3}}  
\frac{(m_{13}-m_{22}+1)(m_{23}-m_{22})
       (m_{33}-m_{22}-1)(m_{11}-m_{22})}
     {(m_{12}-m_{22}+1)(m_{12}-m_{22})}
\right)^{1/2} .                                         \nonumber 
\end{gather}
Action formulas for the operators $E_1^{\pm}$ and 
$H_1$ are completely analogous to formulas (25) above  
(with account of  $m_{11}-m_{22}= 2 j ,\ 
                     2m_{11}-m_{12}-m_{22}= 2 m$).

The presented action formulas for the operators that form the
operator algebra of the $\chi$-type show that their 
matrix elements are, to some extent, similar to 
the 'classical' matrix elements (i.e. to the matrix 
elements of the irrep $\chi$ operators for $su(n)$).
However, there is an essential distinction: now we 
observe the important phase factors (namely, 
$(-1)^{m_{11}+m_{13}+m_{23}+m_{33}}$ under the 
square root in $a_{\chi,m_{11}}$ and $b_{\chi,m_{11}}$)
which {\em depend on} $\chi$ {\em and a specified basis element}.
No such basis-element dependent factors 
exist in the $su(n)$ case.

Let us illustrate such treatment with the particular 
example of operator algebra appearing in the 
singular $q=-1$ case of $U_q(su_3)$ and 
corresponding to the octet representation of $su_3$.
We give here explicitly only those action formulas 
for $E_1^\pm$ and $E_2^\pm$ in which matrix elements 
differ from their corresponding 'classical' counterparts:
\begin{gather}
E_1^-|\Sigma^+\rangle=\sqrt{-2}|\Sigma^0\rangle , \hspace{8mm} 
E_1^-|\Sigma^0\rangle=\sqrt{-2}|\Sigma^-\rangle , \hspace{8mm} 
E_1^+|\Sigma^-\rangle=\sqrt{-2}|\Sigma^0\rangle ,  \nonumber\\
E_1^+|\Sigma^0\rangle=\sqrt{-2}|\Sigma^+\rangle ,      
                                            \hspace{8mm} 
E_2^-|n\rangle=\frac{1}{\sqrt{-2}}|\Sigma^0\rangle+
\sqrt{-3/2}|\Lambda\rangle,                 \hspace{8mm} 
E_2^-|\Lambda\rangle=\sqrt{-3/2} 
           |\Xi^0\rangle,\ \ \ \                   \nonumber
\end{gather}
\begin{gather}
E_2^-|\Sigma^0\rangle=\frac{1}{\sqrt{-2}}|\Xi^0\rangle\ ,  
                                            \hspace{8mm}
E_2^+|\Xi^0\rangle=\frac{1}{\sqrt{-2}}|\Sigma^0\rangle+
\sqrt{-3/2} |\Lambda\rangle,                \hspace{8mm}
E_2^+|\Lambda\rangle=\sqrt{-3/2} 
                                |n\rangle,            \nonumber\\
E_2^+|\Sigma^0\rangle=\frac{1}{\sqrt{-2}}|n\rangle\ .    \nonumber
\end{gather}
To complete this operator algebra, we must add the rest of action formulas
for $E_1^\pm$ and $E_2^\pm$ (i.e., action on those basis elements)
which coincide with the 'classical' ones,
as well as the action formulas for $H_1$, $H_2$ (these latter also
coincide with 'classical' formulas).

Likewise, for $U_q(su_3)$ at $q=-1$ an operator algebra 
corresponding to any other irrep of $su_3$ can be given. 
The treatment is obviously extendible to $U_{q=-1}(su_n),\ n>3$.

Let us also remark that SUSY-based mass relation $m_{\rho}=m_{\omega}$,
see (2), is obtainable from a $q$-deformed structure. 
Indeed, it follows from the $q$-analog of vector meson mass relation,               
\[
m_{{\omega}_8}+(2[2]_q/[3]_q -1) m_{\rho} = (2[2]_q/[3]_q) m_{K^*} 
\]
(which was derived                \cite{gavr:jpha} 
using $U_q(su_n)$ quantum algebras), if one fixes $q$ 
as 4th root of unity: $q=\sqrt{-1}\ $ (then, $[2]_q=0$). 
The intriguing interplay between SUSY and the special 
cases $q=-{1}$ and $q=\sqrt{-1}$ of the $q$-algebras 
$U_q(su_n)$ deserves further detailed study.

\subsection*{Acknowledgement}

The research described in this paper was made possible in part by
Award No. UP1-2115 of the U.S. Civilian Research and Development 
Foundation for Independent States of the Former Soviet Union (CRDF).

\LastPageEnding

\end{document}